\documentclass[a4paper]{jpconf}
\usepackage{graphicx}
\usepackage{bm}

\begin{document}
\title{Neutrino electromagnetic properties and new bounds on neutrino magnetic moments}
\author{K A Kouzakov$^{1}$, A I Studenikin$^{2,3}$ and M B Voloshin$^{4,5}$}
\address{$^1$Department of Nuclear Physics and Quantum Theory of Collisions, Faculty of Physics, Moscow State University, Moscow 119991, Russia}
%

%
\address{$^{2}$Department of Theoretical Physics, Faculty of Physics, Moscow State University, Moscow 119991,
Russia\\
$^{3}$Joint Institute for Nuclear Research, Dubna 141980, Moscow
Region,
Russia}
\address{$^{4}$William I. Fine Theoretical Physics Institute,
University of Minnesota, Minneapolis, Minnesota 55455, USA\\
$^{5}$Institute of Theoretical and Experimental Physics, Moscow
117218, Russia}%

%
 \ead{studenik@srd.sinp.msu.ru}
\begin{abstract}
We give a brief outline of possible neutrino electromagnetic
characteristics, which can indicate new physics beyond the
Standard Model. Special emphasis is put on recent theoretical
development in searches for neutrino magnetic moments.
\end{abstract}
\section{Introduction}
\label{intro}
In particle physics, the neutrino plays a remarkable role of a
``tiny'' particle. Indeed, the scale of neutrino mass $m_\nu$ is
much lower than that of the charged fermions ($m_{\nu_f}\ll m_f$,
$f = e, \mu, \tau$). Interaction of neutrinos with matter is
extremely weak as compared to that in the case of other known
elementary fermions, and it can be mediated via the weak or
electromagnetic channel. In this context, neutrino electromagnetic
properties are of particular interest, for they open a door to
``new physics'' beyond the Standard Model
(SM)~\cite{studenikin09}. In spite of appreciable efforts in
searches for electromagnetic properties of neutrino, up to now
there is no experimental evidence favoring their nonvanishing
electromagnetic characteristics. However, the recent development
of our knowledge of neutrino mixing and oscillations, supported by
the discovery of flavor conversions of neutrinos from different
sources, makes quite plausible the assumption that neutrinos have
``nonzero'' electromagnetic properties. The latter include, in
particular, the electric charge, the charge radius, the anapole
moment, and the dipole electric and magnetic moments.

The neutrino magnetic moments (NMM) expected in the SM are very
small and proportional to the neutrino masses\footnote{ The units
$\hbar=c=1$ are used throughout unless otherwise
stated.}~\cite{fs}:
\begin{equation}
 \mu_\nu \approx 3 \times 10^{-19} \mu_B \left(\frac{m_\nu}{1
\, {\rm eV}}\right), \label{mu_nu}
\end{equation}
with $\mu_B = e/2m$ being the electron Bohr magneton, and $m$ is
the electron mass. Thus any larger value of $\mu_\nu$ can arise
only from physics beyond the SM (a recent review of this subject
can be found in~\cite{gs}). Current direct experimental
searches~\cite{tx,ge1,ge2,ge3} for a magnetic moment of the
electron (anti)neutrinos from reactors have lowered the upper
limit on $\mu_\nu$ down to $\mu_\nu < 3.2 \times 10^{-11} \,
\mu_B$~\cite{ge2,ge3}. These ultra low background experiments use
germanium crystal detectors exposed to the neutrino flux from a
reactor and measure the energy $T$ deposited by the neutrino
scattering in the detector. The sensitivity of such a search to
NMM crucially depends on lowering the threshold for the energy
transfer $T$, due to the enhancement of the magnetic scattering
relative to the standard electroweak one at low $T$.

The paper is organized as follows. In section~\ref{el-m_prop}, we
discuss electromagnetic characteristics that one may expect in the
cases of Dirac and Majorana neutrinos. Specific theoretical
aspects of searches for NMM are considered in
section~\ref{moment}. Section~\ref{conclusions} summarizes this
work.

\section{Electromagnetic properties of neutrino}
\label{el-m_prop}
In general the matrix element of the electromagnetic current
$J_\mu^{\rm EM}$ can be considered between different neutrino
initial $\psi_i(p)$ and final $\psi_j(p')$ states of different
masses, $p^2=m^2_i$ and $p'^2=m^2_j$:
\begin{equation}
\label{matrix_el}\langle\psi_j(p')|J_\mu^{\rm
EM}|\psi_i(p)\rangle=\bar{u}_j(p')\Lambda_\mu(q)u_i(p).
\end{equation}
In the most general case consistent with Lorentz and
electromagnetic gauge invariance, the vertex function is defined
as (see Ref.~\cite{gs} and references therein)
\begin{equation}
\label{vertex}\Lambda_\mu(q)=[f_Q(q^2)_{ij} +
f_A(q^2)_{ij}\gamma_5](q^2\gamma_\mu-\gamma_\mu\!\!\not{q})+f_M(q^2)_{ij}i\sigma_{\mu\nu}q^\nu+f_E(q^2)_{ij}i\sigma_{\mu\nu}q^\nu\gamma_5,
\end{equation}
where $f_Q(q^2)$, $f_A(q^2)$, $f_M(q^2)$, and $f_E(q^2)$ are
respectively the charge, anapole, dipole magnetic, and dipole
electric neutrino form factors, which are matrices in the space of
neutrino mass eigenstates~\cite{shrock82}.

Let us briefly discuss the diagonal case $i=j$. The hermiticity of
the electromagnetic current and the assumption of its invariance
under discrete symmetries transformations put certain constraints
on the neutrino form factors, which are in general different for
the Dirac and Majorana cases. In the case of Dirac neutrinos, the
assumption of CP invariance combined with the hermiticity of the
electromagnetic current $J_\mu^{\rm EM}$ implies that the electric
dipole form factor vanishes, $f_E=0$. At zero momentum transfer
only $f_Q(0)$ and $f_M(0)$, which are called the electric charge
and the magnetic moment, respectively, contribute to the
Hamiltonian $H_{\rm int}\sim J^{\rm EM}_\mu A^\mu$, which
describes the neutrino interaction with the external
electromagnetic field $A^\mu$. Hermiticity also implies that
$f_Q$, $f_A$, and $f_M$ are real. In contrast, in the case of
Majorana neutrinos, regardless of whether CP invariance is
violated or not, the charge, dipole magnetic and electric moments
vanish, $f_Q=f_M=f_E=0$, so that only the anapole moment can be
non-vanishing among the electromagnetic moments. Note that it is
possible to prove~\cite{kayser82,kayser84,nieves82} that the
existence of a non-vanishing magnetic moment for a Majorana
neutrino would bring about a clear evidence for CPT violation.

In the off-diagonal case $i\neq j$ the hermiticity by itself does
not imply restrictions on the form factors of Dirac neutrinos. It
is possible to show~\cite{nieves82} that if the assumption of CP
invariance is added, the form factors $f_Q$, $f_M$, $f_E$, and
$f_A$ should have the same complex phase. For the Majorana
neutrino, if CP invariance holds, there could be either a
transition magnetic or a transition electric moment. Finally, as
in the diagonal case, the anapole form factor of a Majorana
neutrino can be nonzero.

\section{Searches for neutrino magnetic moments}
\label{moment}
The neutrino dipole magnetic and electric form factors (and the
corresponding magnetic and electric dipole moments) are
theoretically the most well studied among the form factors. They
also attract a notable attention from experimentalists, although
the NMM value~(\ref{mu_nu}) predicted in the SM is many orders of
magnitude smaller than the present experimental limits achievable
in terrestrial experiments. The most sensitive and established
method for the experimental investigation of the NMM is provided
by direct laboratory measurements of electron
(anti)neutrino-electron scattering at low energies in solar,
accelerator, and reactor experiments. A detailed description of
various experiments can be found in~\cite{ge1,wong05}.

The cross section for electron (anti)neutrino scattering on a free
electron can be written~\cite{vogel89} (see
also~\cite{ge1,wong05}) as a sum of the SM and NMM contributions,
\begin{equation}
\label{cr_sec}\frac{d\sigma}{dT}=\frac{d\sigma_{\rm
SM}}{dT}+\frac{d\sigma_{(\mu)}}{dT},
\end{equation}
where $E_\nu$ is the incident neutrino energy and $T$ is the
energy transfer. The SM contribution is constant in $T$ at $E_\nu
\gg T$:
\begin{equation} \frac{d\sigma_{\rm SM}}{dT}= {G_F^2  m \over 2 \pi} \left( 1+ 4 \sin^2
\theta_W + 8 \sin^4 \theta_W \right) \left [ 1 + O \left ( {T
\over E_\nu} \right) \right ] \approx  10^{-47} {\rm cm^2/keV}.
\label{SM}
\end{equation}
In contrast, the NMM contribution
\begin{equation} \frac{d\sigma_{(\mu)}}{dT}= 4 \pi  \alpha  \mu_\nu^2
 \left ( {1 \over T} - {1 \over E_\nu } \right ) = \pi {\alpha^2 \over m^2} \left ( {\mu_\nu \over \mu_B} \right )^2
\left ( {1 \over T} - {1 \over E_\nu } \right ) \label{fe}
\end{equation}
exhibits a $1/T$ enhancement at low energy transfer. Note that the
NMM contribution to the cross section changes the helicity of the
neutrino, contrary to the SM contribution and to the possible
contribution from the neutrino charge radius. Therefore, for
relativistic neutrino energies the interference between
$d\sigma_{\rm SM}/dT$ and $d\sigma_{(\mu)}/dT$ is a negligible
effect in the total cross section~(\ref{cr_sec}).

The current experiments with reactor (anti)neutrinos have reached
threshold values of $T$ as low as few keV and are likely to
further improve the sensitivity to low energy deposition in the
detector. At low energies however one can expect a modification of
the free-electron formulas~(\ref{SM}) and~(\ref{fe}) due to the
binding of electrons in the germanium atoms, where e.g. the energy
of the $K_\alpha$ line, 9.89\,keV, indicates that at least some of
the atomic binding energies are comparable to the already relevant
to the experiment values of $T$. In the case $E_\nu\gg T$, which
is relevant to the experiments with reactor (anti)neutrinos, it
can be shown~\cite{voloshin10,kouzakov11,kouzakov11_jetp} that the
SM and NMM contributions to the neutrino scattering on atomic
electrons are
\begin{equation} \frac{d\sigma_{\rm
SM}}{dT}=\left(\frac{d\sigma_{\rm SM}}{dT}\right)_{\rm
FE}\frac{I_1(T)}{2m}, \qquad I_1(T)=\int^\infty_0 S(T,q^2)dq^2,
\label{SM_bind}
\end{equation}
\begin{equation} \frac{d\sigma_{(\mu)}}{dT}=\left(\frac{d\sigma_{(\mu)}}{dT}\right)_{\rm
FE}TI_2(T), \qquad I_2(T)=\int^\infty_0 S(T,q^2)\frac{dq^2}{q^2},
\label{fe_bind}
\end{equation}
where $(d\sigma_{\rm SM}/dT)_{\rm FE}$ and
$(d\sigma_{(\mu)}/dT)_{\rm FE}$ are the free-electron results
given by (\ref{SM}) and (\ref{fe}), respectively. A key quantity
that determines cross sections~(\ref{SM_bind}) and~(\ref{fe_bind})
is the so-called dynamical structure factor $S(T,q^2)$, which is a
function of the energy and momentum transfer values, $T$ and
$q=|{\bf q}|$. For a free electron, one has in a nonrelativistic
limit $S(T,q^2)=\delta(T-q^2/2m)$, which upon substitution
in~(\ref{SM_bind}) and~(\ref{fe_bind}) immediately yields the
free-electron formulas (\ref{SM}) and (\ref{fe}).

Recently it was claimed~\cite{wong10} that the atomic binding
effects must result in a significant enhancement of the NMM
contribution. However, that early claim was later
disproved~\cite{voloshin10,kouzakov11_plb,kouzakov11_npbps}, thus
also disproving the upper bound on the $\mu_\nu$ value,
$\mu_\nu<1.3\times10^{-11}\mu_B$, obtained in~\cite{wong10}. It
was demonstrated~\cite{kouzakov11,kouzakov11_jetp} by means of
analytical and numerical calculations that the atomic binding
effects are adequately described by the so-called stepping
approximation introduced in~\cite{kopeikin97} from interpretation
of numerical data. According to the stepping approach, the SM and
NMM contributions are simply given by
\begin{equation}
\label{step}\frac{d\sigma_{\rm SM}}{dT}=\left(\frac{d\sigma_{\rm
SM}}{dT}\right)_{\rm FE}\sum_in_i\theta(T-E_i), \qquad
\frac{d\sigma_{(\mu)}}{dT}=\left(\frac{d\sigma_{(\mu)}}{dT}\right)_{\rm
FE}\sum_in_i\theta(T-E_i),
\end{equation}
where the $i$ sum runs over all occupied atomic sublevels, with
$n_i$ and $E_i$ being their occupations and binding energies. The
following important conclusions can be drawn from the stepping
approximation~(\ref{step}). Firstly, the atomic effects reduce the
SM and NMM contributions compared to their free-electron values.
Secondly, the ratio between the SM and NMM contributions is not
affected by the atomic binding effects.

\section{Conclusion}
\label{conclusions}
The above theoretical findings strongly support the upper limit
$\mu_\nu < 3.2 \times 10^{-11}\mu_B$ recently reported by the
GEMMA collaboration~\cite{ge2,ge3}. This bound obtained in
terrestrial experiments with reactor (anti)neutrinos is only by an
order of magnitude weaker than the most stringent astrophysical
constraint $\mu_\nu < 3 \times 10^{-12}\mu_B$~\cite{raffelt90}. A
general and model-independent upper bound on the Dirac NMM, that
can be generated by an effective theory beyond the SM, is
$\mu_\nu\leq10^{-14}\mu_B$~\cite{bell05} (the limit in the
Majorana case is much weaker). Thus, the searches for NMM are
close to the territory where new physics can reveal itself.

\ack
One of the authors (A.I.S.) is thankful to Lothar Oberauer, Georg
Raffelt and Robert Wagner for the kind invitation to participate
in the 12th International Conference on Topics in Astroparticle
and Underground Physics. The work was supported by RFBR grant no.
11-02-01509-a.

\section*{References}

\end{document}